\documentclass[UTF8]{elsarticle}

\newtheorem{remark}{Remark}
\usepackage{algorithm}
\usepackage{algorithmic}
\usepackage{slashbox}
\usepackage{graphicx,natbib,amssymb,lineno,color}
\usepackage{url}
\usepackage{amsmath}
\usepackage{setspace}
\usepackage{subfigure}

\usepackage{subfigure}
\usepackage{color}

\journal{Journal of \LaTeX\ Templates}

\begin{document}
\begin{frontmatter}

\title{Privacy-Preserving Cloud-Aided Broad Learning System}

\author[mymainaddress]{Haiyang Liu}

\author[mymainaddress]{Hanlin Zhang\corref{mycorrespondingauthor}}
\cortext[mycorrespondingauthor]{Corresponding author}
\ead{hanlin@qdu.edu.cn}

\author[mymainaddress]{Li Guo}
\author[mymainaddress,mysecondaryaddress,mythirdaddress]{Jia Yu}
\author[myforthaddress]{Jie Lin}

\address[mymainaddress]{College of Computer Science and Technology, Qingdao University\\
266071 Qingdao, China}
\address[mysecondaryaddress]{Institute of Big Data Technology and Smart City, Qingdao University\\
266071 Qingdao, China}
\address[mythirdaddress]{State Key Laboratory of Information Security, Institute of Information Engineering,\\ Chinese Academy of Sciences, 100093 Beijing, China}
\address[myforthaddress]{School of Electronic and Information Engineering, Xian Jiaotong University\\
710049 Xian, China}

\begin{abstract}
With the rapid development of artificial intelligence and the advent of the 5G era, deep learning has received extensive attention from researchers. Broad Learning System (BLS) is a new deep learning model proposed recently, which shows its effectiveness in many fields, such as image recognition and fault detection. However, the training process still requires vast computations, and therefore cannot be accomplished by some resource-constrained devices. To solve this problem, the resource-constrained device can outsource the BLS algorithm to cloud servers. Nevertheless, some security challenges also follow with the use of cloud computing, including the privacy of the data and the correctness of returned results. In this paper, we propose a secure, efficient, and verifiable outsourcing algorithm for BLS. This algorithm not only improves the efficiency of the algorithm on the client but also ensures that the clients sensitive information is not leaked to the cloud server. In addition, in our algorithm, the client can verify the correctness of returned results with a probability of almost 1. Finally, we analyze the security and efficiency of our algorithm in theory and prove our algorithms feasibility through experiments.
\end{abstract}

\begin{keyword}
\texttt{broad learning system (BLS), deep learning, secure outsourcing computations, privacy preserving.}
\end{keyword}

\end{frontmatter}

\section{Introduction}
\begin{spacing}{1.2}
Recently, deep neural network has become a research hotspot in the fields of artificial intelligence and big data analysis, which demonstrates impressive performance, especially in fault detection, face recognition \cite{G1} and spam detection \cite{H2}. Although the deep neural network has many advantages, the deep neural network's training process is time-consuming because of the characteristics of its multilayer network structure. In addition, when the model fails to achieve the expected performance, the time-consuming training process needs to be repeated many times. To address this issue, Broad Learning System (BLS) was proposed by Chen \cite{C6}. The BLS is a new neural network based on the random vector functional link neural network (RVFLNN). Instead of retraining the model from scratch, BLS can reconstruct the existing neural network quickly by incremental learning. Nowadays, BLS demonstrates efficient learning capability in a wide range of applications, e.g., fault diagnosis \cite{H7} and image recognition \cite{J8}. Nevertheless, with the increase of data sets and nodes in the network, the scale of the matrix that BLS needs to deal with in the training process can reach tens of thousands or even millions. Although compared with other deep structures, the training speed of BLS has been greatly improved, it still contains complex calculations.

Due to the limited computing power, some end devices may not be able to complete the training process of BLS. The wireless intelligent sensor camera in the Internet of Things is a typical end device with limited computing power. In some cases, if these wireless intelligent sensors with limited computing power can train the BLS model and execute the next step based on the results, it will greatly reduce the reaction time of IoT devices and improve the efficiency of data analysis and processing. For example, in the intelligent transportation system, if the intelligent wireless sensor device can identify the vehicles on the road, it can timely adjust the signal lights according to the road conditions, so as to effectively relieve the traffic pressure. Due to the limitation of their own structure, the computing power of these wireless sensor devices is very limited \cite{M999}. Therefore, most of them can only be used for information acquisition, not for training model.

The emergence of cloud computing provides a solution to the above problem. Cloud computing is an Internet-based computing resource interaction mode, which integrates various services, applications and other resources, and can be provided to client in the form of services through the Internet. Cloud computing can quickly deploy resources or obtain services based on virtualization technology. At the same time, cloud computing services can provide resources on demand according to client requirements to achieve dynamic resource expansion, and pay according to usage, so that client can conveniently use cloud computing services and reduce The processing burden of the client. Benefiting from the above-mentioned characteristics of cloud computing, the client can use a variety of end devices to access resources in the cloud from any network coverage area without worrying about the amount of resources and capacity planning. This allows resource-constrained client uses cloud computing to solves heavy computing tasks.

Although cloud computing has many advantages, its unique service outsourcing and other characteristics also bring unprecedented security challenges. First, the cloud may disclose the client's sensitive information, which will cause disastrous consequences to the client. Second, client may receive an incorrect result. The incorrect results can be caused by software errors and malicious external attacks. Also, the cloud server may return a random result to the client in order to save computing resources. To solve these problems, the following requirements for outsourcing computations should be met:
$(1)$ The designed algorithm must ensure the confidentiality of the data, which means that the cloud server cannot obtain the client's sensitive information. $(2)$ Client can verify the calculation results returned by the cloud server. $(3)$ The designed algorithm must ensure that the amount of work (including transformation, recovery, and correctness verification) required by the client under this algorithm is less than the amount of work to perform the original computation on its own. Otherwise, the client does not need to seek the help of the cloud. From the perspective of applications, an outsourcing computing algorithm should be secure, verifiable, and efficient.

In this paper, we propose a secure and efficient outsourcing algorithm for BLS. Our algorithm solves the problem that the resource constrained end devices can't train the BLS model. We verify the feasibility of our algorithm through specific experiments.
The main contributions of this paper are summarized as follows:
\begin{itemize}
\item As far as we know, our proposed algorithm is the first secure outsourcing algorithm for the Broad Learning System. Our algorithm allows the client to use the cloud server to safely and effectively solve the high-dimensional data problem in the BLS, so that the BLS can be widely used in some resource-constrained devices.
\item In our algorithm, all inputs and outputs are obscured, and the cloud cannot obtain any sensitive information. Also, the client can effectively verify the results of cloud server calculations.
\item Our algorithm can greatly reduce the training time of BLS and save local computing resources. We prove the correctness, privacy, reliability and efficiency of our proposed algorithm from theoretical and experimental perspectives. In our algorithm, the client computation complexity is reduced from $O(n^{3})$ to $O(n^{2})$.
\end{itemize}

The general structure of this paper is as follows. In section II, we introduced broad learning system and outsourced computing. In section III, we describe the system model and design goals. We introduce the process of outsourcing algorithm in section IV. In section V, we evaluate the algorithm through perfect experiments. In section VI, we briefly introduce the related research on BLS and security outsourcing. Finally, we give a conclusion in section VII.

\section{Research background}
In this section, we first give a brief analysis and introduction of BLS. Also, a simple introduction and analysis to cloud computing is given.
\subsection{Broad Learning System}
Broad Learning System (BLS)  is a type of neural network based on RVFLNN \cite{Pao1992Functional, Pao1994Learning, Igelnik1995Stochastic} proposed by Professor Chen. Different from the traditional deep structure, BLS optimizes the network structure by extending the network width. In BLS, the original input is transmitted by building feature nodes, and the structure is extended by building enhanced nodes. All mapped feature nodes and enhancement nodes can be directly connected to the output, and obtain the expected connection weight through the pseudoinverse. In addition, when the existing model can not meet the client's requirements, BLS can make the model more perfect by incremental learning, and the tedious retraining process is saved. The detailed process of the BLS algorithm is as follows:
\begin{figure}[htb]
\begin{center}
\includegraphics[height=230pt,width=280pt]{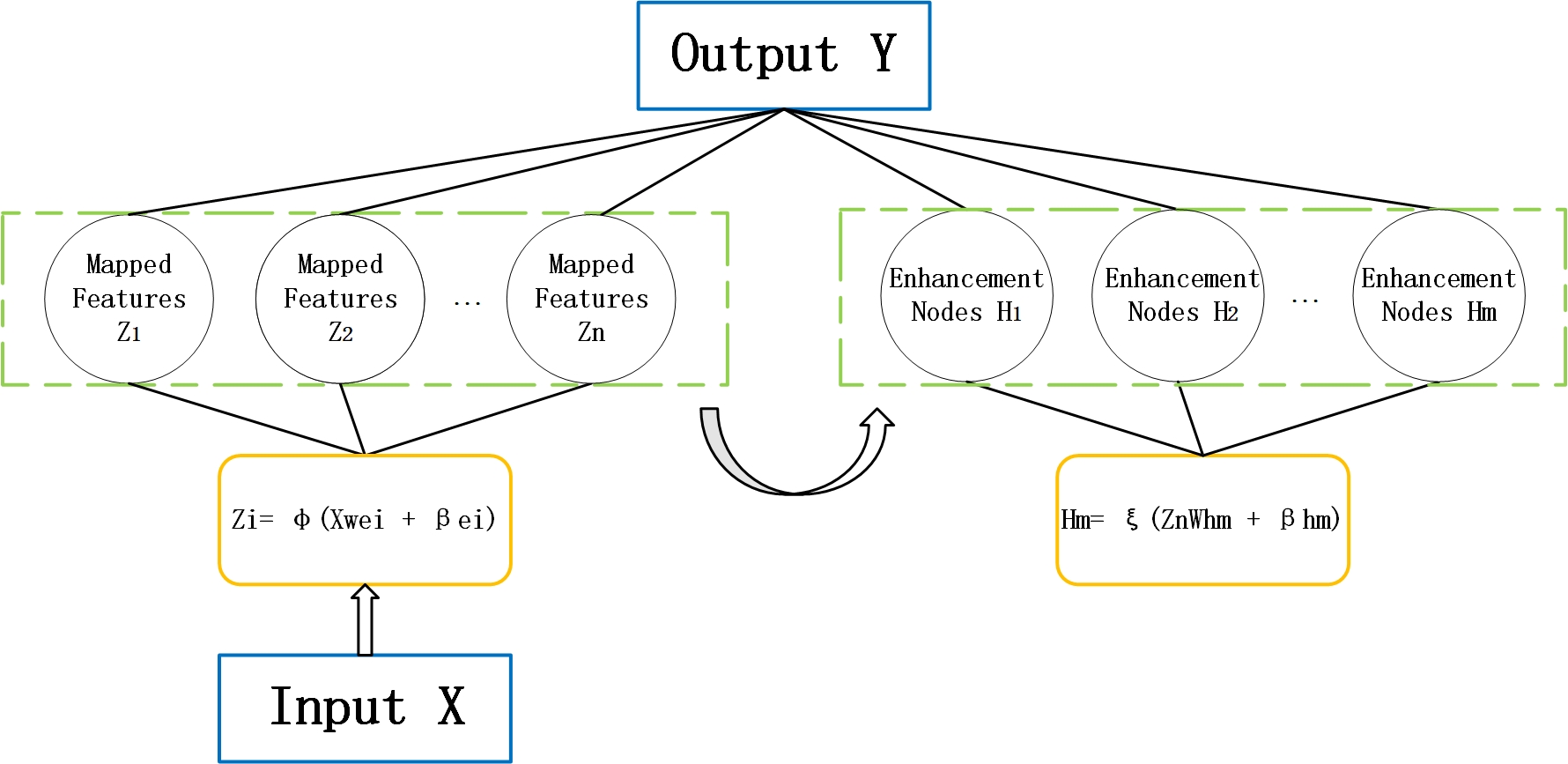}\\
\vspace{-3mm}
\begin{center}
\caption{BLS model}
\label{secmodel} \vspace{-3mm}
\end{center}
\end{center}
\end{figure}

\begin{enumerate}[a)]
\item Select $n$ samples as the training set, where each sample has $m$ dimensions. Before training, we need to perform preprocessing operations such as geometric normalization and image normalization on the data, all data is arranged to form a matrix $X^{n\times m}$.
\item For $n$ groups of feature mapping nodes, the generation method of each group of feature mapping nodes is as follows.
\begin{gather} \label{eq1}
   Z_{i} = \phi(XW_{fi}+ \beta_{fi}),    i = 1, . . . ,n,
\end{gather}
where $W_{fi}$ and $\beta_{fi}$ are randomly generated.All feature nodes are represented as $Z^{n}=[Z_{1},. . . ,Z_{n}]$.
\item Similarly, for $m$ groups of enhancement nodes, the generation method of each node  is as follows.
\begin{gather} \label{eq2}
   H_{j}\equiv \xi(Z_{n}W_{hj} + \beta_{hj}), i = 1, . . . ,m,
\end{gather}
where $W_{hj}$ and $\beta_{hj}$ are randomly generated. All enhancement nodes are represented as $H^{m}=[H_{1},. . . ,H_{m}]$.
\item Join the matrices $Z^{n}$ and $H^{m}$ horizontally to form a matrix $A$.
\item Therefore, the BLS model can be expressed by the following equation
\begin{gather}\label{eq3}
\begin{split}
   Y &= [Z_{1},..., Z_{n}|H_{1},..., H_{m}]W^{m}\\
     &= [Z^{n}|H^{m}]W^{m}\\
     &= AW^{m},
\end{split}
\end{gather}
where Y is the output matrix which belongs to $R^{n\times c}$, $W^{m}= [Z^{n}|H^{m}]^{+}Y=A^{+}W^{m}$. $W^{m}$ are the connecting weights of BLS, which can be approximately calculated through pseudoinverse.
\item Select some samples as the testing set and repeat steps $a$ to $d$. The prediction result can be obtained by multiplying the matrix $A_{1}$ extracted from the testing set with the parameter $W^{m}$.
\end{enumerate}

\begin{remark}
 Pseudoinverse is considered to be a very convenient method to solve the output layer of neural network, which can be easily computed through the ridge regression approximation. Ridge regression can be represented as follow.

\begin{gather}\label{eq4}
  \arg\min _{W}: \|AW-Y\|^{\sigma_{1}}_{\nu}+\lambda\|W\|^{\sigma_{2}}_{\mu},
\end{gather}
where $ \sigma_{1}= \sigma_{2}= \nu = \mu= 2$. $\lambda$ is the coefficient of the sum of squares of the weight W, which represents the constraint on the sum of squares of the weight. This solution is equivalent with the ridge regression theory, which adding a positive number to the diagonal of $A^{T}A$ or $AA^{T}$ to approximate Moore-Penrose generalized inverse \cite{Hoerl2000Ridge}. Theoretically, if $\lambda= 0$, the inverse problem degenerates into the least square problem. However, if $\lambda \rightarrow\infty$, the solution is heavily constrained and approaches to 0. Accordingly, we have
\begin{gather}\label{eq5}
   W = (\lambda I + A^{T}A)^{-1}A^{T}Y,
\end{gather}

where $I$ is the unit matrix. Specifically, we have that

\begin{gather}\label{eq6}
     A^{+} = \lim_{\lambda\rightarrow0}(\lambda I + A^{T}A)^{-1}A^{T}.
\end{gather}

Derive from the above, we can obtain the pseudoinverse matrix we want, and the parameter $W$ can be easily obtained by using the pseudoinverse matrix $A^{+}$.
\end{remark}

\subsection{Cloud Computing}

Cloud computing is a type of distributed computing, which provides available, convenient and on-demand network access. With cloud computing technology, client can use the cloud server to complete complex calculations and massive data processing, which can significantly reduce the local computation overhead. The powerful storage capacity of the cloud server enables different end devices to realize data sharing when using cloud computing services. As long as the client can access the Internet, they can access the information stored in the cloud and applications on the cloud. At the same time, the price of cloud computing services is low. The client can flexibly use various services provided by the cloud server according to their own computing needs. Cloud is a huge resource pool, which gathers a lot of computing resources. And cloud computing center can integrate and manage these huge computing resources to give client unprecedented computing and storage capabilities. In short, the core concept of cloud computing is to take the Internet as the center, and connect the tangible and intangible resources associated with the Internet to form a platform to provide the client with fast and safe computing and storage services, so that everyone on the Internet can use the powerful computing resources and storage resources in the cloud server.

\section{System Model and Design Goals} \label{sec:rel}
In this section, we will introduce the system model, design objectives of  outsourcing computation.

\subsection{System Model}

\begin{figure}[htb]
\begin{center}
\includegraphics[height=150pt,width=250pt]{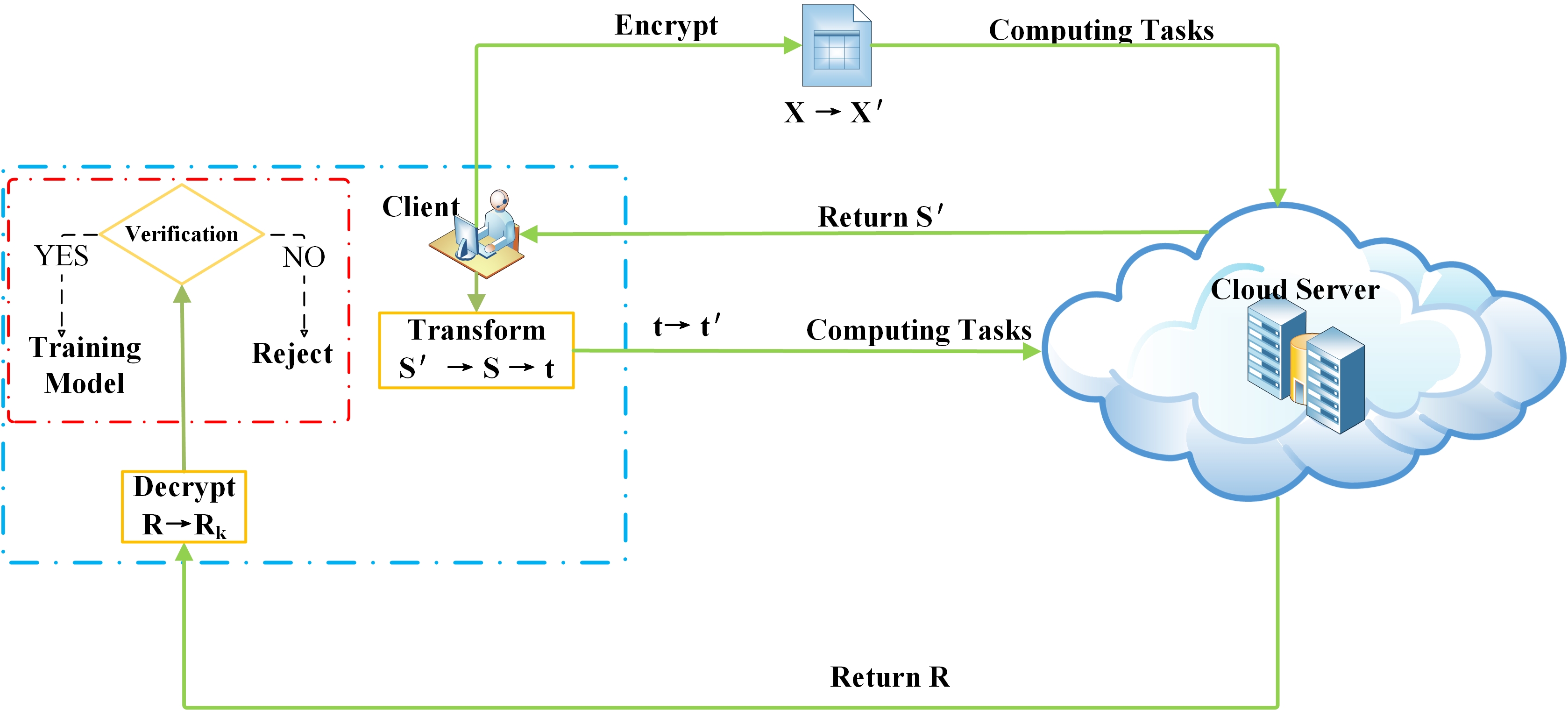}\\
\vspace{-3mm}
\begin{center}
\caption{System model}
\label{System Model} \vspace{-3mm}
\end{center}
\end{center}
\end{figure}

As shown in Fig. \ref{System Model}, assume that the computing power and memory of the client are severely limited, and the client needs to train the BLS model \cite{Shoubo2019Robust, Tong2019Facial}. Due to the limited local computing power, the training process of BLS cannot be completed. Therefore, the client leverages the powerful computing power of cloud servers to solve this problem. However, the cloud server can not be fully trusted as it is generally controlled by private enterprises. Therefore, the client needs to ensure that any private data in the input and output cannot be leaked. The client first generates the key matrix $K$ and stores it locally, and then uses the key matrix $K$ to obscure the input information $A$. Next, the client sends the computing task $f(x)$ and input $A'$ to the cloud server. The cloud server performs calculation tasks to obtain the calculation result $S$, and returns the result $S$ to the client. Then, the client recovers $S$, and calculate a value $t$ locally based on the recovered result. After that, the client encrypts $t$ as $t'$ with the key matrix. The client sends the next computing task $g(x)$ and $t'$ to the cloud server. The cloud server executes the calculation task $g(x)$ and returns the calculation result $R$ to the client. Next, the client recovers $R_{k}$ from the result $R$. Finally, the client verifies the result. If the verification passes, the client accepts the result; otherwise, if the client detects an error or fraud, the client rejects $R_{k}$.

\subsection{Threat Model and Design Goals}
In our paper, we consider the cloud server to be malicious, which means that the cloud server may steal the client's privacy information. Also, the cloud may return an invalid result to the client. In order to ensure that client can use cloud computing services safely, our algorithm should meet the following goals.
\begin{itemize}
\item \textbf{Correctness:} If the outsourcing algorithm is executed honestly, the final result obtained by the client must be correct.
\item \textbf{Security:}  During the execution of the algorithm by the cloud server, the server cannot obtain any private data about the client.
\item \textbf{Soundness:} The algorithm must ensure that client will not receive the incorrect results.
\item \textbf{Efficiency:} In the outsourcing algorithm, the amount of local calculation performed by the client should be sufficiently less than the amount of calculation required by the client to solve the original problem on its own.
\end{itemize}

\section{Secure Outsourcing Algorithm for Broad Learning System} \label{sec:rel}
In this section, we first explain the basic idea of how we design secure outsourcing algorithms for the broad learning system. Then, we introduce the framework and details of our proposed outsourcing algorithm.

\subsection{Design Rationale}
Our goal is to design a secure and efficient outsourcing algorithm for broad learning system, in which all input and output will be obscured. Our outsourcing algorithm enables the client to safely outsource complex computing to the cloud server. In BLS, using ridge regression to find pseudoinverse is the most time-consuming part. The specific process of calculating pseudoinverse is as follows:
\begin{gather}\label{eq7}
    A^{+} = \lim_{\lambda\rightarrow0}(\lambda I + A^{T}A)^{-1}A^{T}.
\end{gather}
Because it involves large-scale matrix operation, the time complexity of this part can reach $O(n^{3})$. Thus, our protocal focuses on improve the efficiency of this process.

The first step in an outsourcing algorithm is to calculate $A^{T}A$. $A^{T}$ and $A$ have to be obscured because they contain sensitive information. Similarly, the calculation result $A^{T}A$ should also be obscured. Next, the cloud server calculates Eq.(\ref{eq7}) and returns the final result $A^{+}$ to the client. In this process, the inputs $(\lambda I + A^{T}A)^{-1}$ and $A^{T}$, the output $A^{+}$ need to be obscured.

Now we roughly describe how to obscure input and output. In the first step of the outsourcing algorithm for BLS, inputs $A$ and $A^{T}$ need to be obscured. In order to reduce the amount of calculation for the client to obscure the input information, we use two special matrices (sparse matrix $Q$ and sparse positive definite matrix $P$) to obscure the inputs. The advantage of using these two special matrices is that after obscuring matrix $A$ to get $A'$, only need to calculate $(A')^{T}$ for $A'$ to get the fuzzy result of matrix $A^{T}$. After the above blinding operations, the input and output are effectively obscured. The client can make full use of the special properties of the orthogonal matrix to easily recover the real results. In the algorithm, the obscuring process of the second step has the same logic as the first step.

\subsection{Generic Framework}
The proposed outsourcing algorithm for BLS should be composed of the following seven sub-algorithms.
\begin{itemize}
\item \textbf{ProbTransformation1:} The client generates keys $K_1$ and $K_2$ randomly and stores it locally. In order to obscure the private information, the client encrypts the original matrix $\Phi$ to $\Phi_{K_1}$ with the key $K_1$.
\item \textbf{Computation1:} The cloud calculates the corresponding result $S$ according to the algorithm and returns it to the client.
\item \textbf{Recovery1:} The client uses the key matrix $K_1$ to recover $S_{K_1}$ from the result $S$ returned by the cloud.
\item \textbf{ProbTransformation2:} After obtaining $S_{K_1}$, the client uses $\lambda$ and $S_{K_1}$ to calculate value $\mathbf{t}$. Next, client uses key matrix $K_2$ to encrypt $t$ to obtain $t'$.
\item \textbf{Computation2:} After receiving the input $t'$, the cloud calculates result $R'$ and returns $R'$ to the client.
\item \textbf{Recovery2:} After receiving $R'$, the client decrypts $R'$ using the key matrix $K_2$ to obtain the real result $R$.

\item \textbf{Verification:} After obtaining the decrypted result $R$, the client verifies that $R$ is correct. If the verification result is correct, The client accepts the result $R$; otherwise, if an error or fraud is detected in the result, the client rejects $R$.
\end{itemize}

\subsection{Algorithm Details}

Algorithm \ref{Algorithm1} shows the process of key matrix construction.
Firstly, we define two key matrices $P$ and $Q$, which are based on permutation and Kronecker delta function. $P$ can be constructed as follows:
\begin{gather}\label{eq8}
  P(i,j) = \omega_{i}\delta_{\pi_{1}(i),j} \hspace{2em}  1 \leqslant i,j \leqslant m,
\end{gather}
where $\omega_{1},. . . , \omega_{m}$ is from random space $\Omega = \{-1,1\}$, $\pi_{1}$ is a random permutation $\pi_{1}\in \{1, \ldots , m\},$ and $\delta_{x,y}$ is Kronecker delta function. The construction of Kronecker delta function is as follows

\begin{equation}\label{eq9}
	\delta_{x,y}=\left\{
		\begin{aligned}
		1&, x = y, \\
		0&, x\neq y.
	\end{aligned}
	\right.
\end{equation}
The key matrix $Q$ can be constructed as below:
 \begin{gather}\label{eq10}
  Q(i,j) = a(i)\delta_{\pi_{2}(i),j} \hspace{2em}  1 \leqslant i,j \leqslant n,
\end{gather}
where $\pi_{2}$ is a random permutation, $\pi_{2}\in \{1, \ldots , n\}$, $a$ is a set of nonzero random integers, $a_{i}\in \{1, \ldots , n\}\hspace{1em} i = 1,2, . . . ,n$. It is easy to find inverse matrix for both $P$ and $Q$, $Q^{-1}(i,j) = \pi_{2}(i)^{-1}\delta_{\pi^{-1}_{2}(i),j}$. Since $P$ is a sparse orthogonal matrix, $P^{-1}=P^{T}$.

\begin{algorithm} \algsetup{linenosize=\small} \small
 \caption{ Constructing Key Matrix}
 \label{Algorithm1}
 \begin{algorithmic}[1]
 \REQUIRE~~\\
  $m$ and $n$.
  \ENSURE~~\\
   $P$ and $Q$.
 \STATE Set two random permutations $\pi_{1}$, $\pi_{2}$,where $\pi_{1}\in \{1, \ldots , m\}$and $\pi_{2}\in \{1, \ldots , n\}.$\\
  \STATE Set a group of random numbers $\omega_{1},. . . , \omega_{m}$ from random space $\Omega = \{-1,1\}.$
  \STATE \textbf{For} each $i$ from 1 to $m$ \textbf{do}\\
  \STATE \hspace{2em} \textbf{For} each $j$ from 1 to $m$ \textbf{do}\\
    \STATE \hspace{2em}\hspace{2em} $P(i,j) = \omega_{i}\delta_{\pi_{1}(i),j}.$\\
    \hspace{2em} \textbf{End For}
  \STATE\textbf{End for}
    \STATE \textbf{For} each $i$ from 1 to $n$ \textbf{do}\\
  \STATE \hspace{2em} \textbf{For} each $j$ from 1 to $n$ \textbf{do}\\
    \STATE \hspace{2em}\hspace{2em} $Q(i,j) = \pi_{2}(i)\delta_{\pi_{2}(i),j}.$\\
    \hspace{2em} \textbf{End For}
  \STATE\textbf{End for}
    \STATE Output $P$ and $Q$.\\
 \end{algorithmic}
\end{algorithm}

The process of obscuring the matrices and outsourcing computation is shown in Algorithm \ref{Algorithm2}.
In the outsourcing algorithm, the client wants to get the pseudoinverse of matrix $A$. Simultaneously, the client expects that neither matrix $A$ nor its pseudoinverse is exposed to the cloud. To obscure the inputs and output, the client obscures $A$ as follows:
\begin{gather}\label{eq11}
  A' = PAQ.
\end{gather}
The matrix $A$ is transformed into matrix $A'$. Next, the client sends $A'$ to the cloud server and calculate ${A'}^{T}A'$
\begin{gather}\label{eq12}
  {A'}^{T}A'=Q^{T}A^{T}P^{T}PAQ.
\end{gather}
Since P is a sparse orthogonal matrix, $P^{T}P=E$. Consequently, we have that
\begin{gather}\label{eq13}
  {A'}^{T}A'=Q^{T}A^{T}AQ.
\end{gather}
Then the cloud returns the calculation result ${A'}^{T}A'$  to the client. From Eq.(\ref{eq13}), we know that the real result is obscured by key matrices $P$ and $Q$, the client decrypts ${A'}^{T}A'$ with the key matrix to get the correct result $A^{T}A$ easily.
\begin{gather}\label{eq14}
\begin{split}
  {A}^{T}A&=({Q^{T}})^{-1}{A'}^{T}A'Q^{-1}\\
          &=({Q^{T}})^{-1}Q^{T}A^{T}AQQ^{-1}
\end{split}
\end{gather}
Next, client uses $A^{T}A$ to calculate $R_{1}$ and obscure the sensitive information:
\begin{gather}
\begin{split}\label{eq15}
    R_{1} &= \lim_{\lambda\rightarrow0}(\lambda I + A^{T}A)\\
    R_{2} &= Q^{T}R_{1}Q.
\end{split}
\end{gather}
The client sends the calculated $R_{2}$ to the cloud server. Then the cloud server calculates the inverse matrix of $R_{2}$.
\begin{gather}\label{eq16}
R_{2}^{-1}= Q^{-1}R_{1}^{-1}(Q^{T})^{-1}.
\end{gather}
Then client commands the cloud server to calculate the product of the inverse matrix of $R_{2}$ and ${A'}^{T}$ as $R_{3}$:
\begin{gather}\label{eq17}
\begin{split}
  R_{3}&=R_{2}^{-1}{A'}^{T}\\
         &=Q^{-1}R_{1}^{-1}(Q^{T})^{-1}Q^{T}A^{T}P^{T}\\
         &=Q^{-1}R_{1}^{-1}A^{T}P^{T}\\
         &=Q^{-1}\lim_{\lambda\rightarrow0}(\lambda I + A^{T}A)^{-1}A^{T}P^{T}.
\end{split}
\end{gather}
As discussed above, the client protects both the input and output privacy through the process of outsourcing.

\begin{algorithm} \algsetup{linenosize=\small} \small
 \caption{Obscure Matrix and Outsourcing Computation}
 \label{Algorithm2}
 \begin{algorithmic}[1]
  \REQUIRE~~\\
  Matrix A.
  \ENSURE~~\\
   $R_{3}$   (Results returned by cloud computing).
  \STATE Client calculates$A' = PAQ$. \\
  \STATE Client sends $A'$ to cloud. \\
  \STATE Client calculates ${A'}^{T}A'$. \\
  \STATE Client returns ${A'}^{T}A'$ to client. \\
  \STATE Client decrypts ${A'}^{T}A'$ with the key matrix. \\
   ${A}^{T}A=({Q^{T}})^{-1}{A'}^{T}A'Q^{-1}.$
  \STATE Client calculate $$ R_{1} = \lim_{\lambda\rightarrow0}(\lambda I + A^{T}A).$$\\
  \STATE Client calculates $R_{2} = Q^{T}R_{1}Q.$ \\
  \STATE Send $R_{2}$ to cloud.\\
  \STATE Could calculates $R_{3} = R_{2}^{-1}{A'}^{T}.$ \\
  \STATE Output $R_{3}.$\\
 \end{algorithmic}
\end{algorithm}

After receiving the result, the client recovers the result $R_{3}$ with key matrices $P$ and $Q$.
As the relationship between real result and $R_{3}$ is showed by Eq.(\ref{eq17}), the client can easily get the real result:
\begin{gather}\label{eq18}
\begin{split}
  R_{4}&=QR_{3}P\\
       &=\lim_{\lambda\rightarrow0}(\lambda I + AA^{T})^{-1}A^{T}.
\end{split}
\end{gather}

\begin{algorithm} \algsetup{linenosize=\small} \small
 \caption{ Verification Algorithm}
 \label{Algorithm3}
 \begin{algorithmic}[1]
   \REQUIRE~~\\
  $m.$
  \ENSURE~~\\
   $R_{4}.$

  \STATE Client randomly selects a $m \times 1$ vector $\gamma$ whose every element is randomly generated.\\
  \STATE Client computes $A \gamma.$\\
  \STATE Client computes $R_{4}(A \gamma).$\\

  \STATE\textbf{If} $R_{4}(A \gamma)\neq \gamma$\\
    \STATE\hspace{2em} Output ``reject the wrong result".\\
  \STATE\textbf{End if}
    \STATE Output $R_{4}.$\\
 \end{algorithmic}
\end{algorithm}
The verification procedure is shown in Algorithm \ref{Algorithm3}. If $R_{4}$ is the pseudoinverse of matrix $A$, then $R_{4}$ should satisfy
\begin{gather}\label{eq19}
R_{4}A=I.
\end{gather}
However, the time complexity of the matrix multiplication on the left side is $O(n^{3})$. To solve this problem, the client needs generates a random vector $\gamma$, then the client verifies whether
\begin{gather}\label{eq20}
R_{4}A \gamma=\gamma.
\end{gather}

To verify if the above equation holds, the client first calculates $(A \gamma)$, in which time complexity is $O(n^{2})$. Next, the client calculates $R_{4}(A \gamma)$. Because $(A \gamma)$ is also a vector, the time complexity is also $O(n^{2})$ to calculate $R_{4}(A \gamma)$.
If the result returned by the cloud passes the verification of algorithm \ref{Algorithm3}, the client accepts the result. Otherwise, the client declares that the result is wrong and asks the cloud to calculate again until the correct result is returned.

The complete secure outsourcing algorithm for BLS is shown in Algorithm \ref{Algorithm4}.
\begin{algorithm} \algsetup{linenosize=\small} \small
 \caption{BLS-based Cloud Computing Outsourcing Algorithm}
 \label{Algorithm4}
 \begin{algorithmic}
   \REQUIRE~~\\
  Matrix A.
  \ENSURE~~\\
   R4 ( Real Result).
  \STATE\textbf{1. } \textit{\textbf{ProbTransformation1}}
  \begin{itemize}
    \STATE Set two random permutations $\pi_{1}$, $\pi_{2}$,where $\pi_{1}\in \{1, \ldots , m\}$ and $\pi_{2}\in \{1, \ldots , n\}.$\\
  \STATE Set a group of random numbers $\omega_{1},. . . , \omega_{m}$ from random space $\Omega = \{-1,1\}.$

  \STATE \textbf{For} each $i$ from 1 to $m$ \textbf{do}\\
   \hspace{2em} \textbf{For} each $j$ from 1 to $m$ \textbf{do}\\
     \hspace{2em}\hspace{2em} $P(i,j) = \omega_{i}\delta_{\pi_{1}(i),j}.$\\
     \hspace{2em} \textbf{End For}\\
  \textbf{End for}
    \STATE \textbf{For} each $i$ from 1 to $n$ \textbf{do}\\
   \hspace{2em} \textbf{For} each $j$ from 1 to $n$ \textbf{do}\\
     \hspace{2em}\hspace{2em} $Q(i,j) = \pi_{2}(i)\delta_{\pi_{2}(i),j}.$\\
     \hspace{2em} \textbf{End For}\\
  \textbf{End for}
  \STATE Client calculates$A' = PAQ.$\\
  \end{itemize}

  \STATE\textbf{2. } \textit{\textbf{Computation1}}
  \begin{itemize}

  \STATE Client sends $A'$ to cloud. \\
  \STATE Cloud calculates ${A'}^{T}A'.$ \\
  \STATE Cloud returns ${A'}^{T}A'$ to client. \\
  \end{itemize}
\STATE\textbf{3. } \textit{\textbf{Recovery1}}
   \begin{itemize}
  \STATE Client decrypts ${A'}^{T}A'$ with the key matrix. \\
   ${A}^{T}A=({Q^{T}})^{-1}{A'}^{T}A'Q^{-1}.$
  \end{itemize}
  \STATE\textbf{4. } \textit{\textbf{ProbTransformation1}}
  \begin{itemize}
  \STATE Client calculates $$ R_{1} = \lim_{\lambda\rightarrow0}(\lambda I + A^{T}A). $$\\

  \STATE Client calculates $R_{2} = Q^{T}R_{1}Q.$ \\
  \end{itemize}

  \STATE\textbf{5. } \textit{\textbf{Computation2}}
  \begin{itemize}
  \STATE Send $R_{2}$ to cloud.\\

   Could calculates $R_{3} = R_{2}^{-1}{A'}^{T}.$ \\
  \end{itemize}

  \STATE\textbf{6. } \textit{\textbf{Recovery2}}
  \begin{itemize}
  \STATE Client decrypts the result $R_{3}$ returned by the cloud with keys matrix $P$ and $Q$. \\
   $$ R_{4}=QR_{3}P=\lim_{\lambda\rightarrow0}(\lambda I + AA^{T})^{-1}A^{T}.$$ \\
 \end{itemize}

\STATE\textbf{7. } \textit{\textbf{Verification}}
\begin{itemize}

  \STATE Client randomly selects an $m \times 1$ vector $\gamma$ whose every element is randomly generated.\\
  \STATE Client computes $A \gamma$. Then the client computes $R_{4}(A \gamma).$\\

  \STATE\textbf{If} $R_{4}(A \gamma)\neq \gamma$\\
    \hspace{2em} Output ``reject the wrong result".\\
     \hspace{2em} End the algorithm.\\
  \textbf{End if}
\end{itemize}
    \STATE Output $R_{4}.$\\
 \end{algorithmic}
\end{algorithm}

\section{Analysis of Outsourcing Algorithm for Broad Learning System}
In this section, we prove that our proposed outsourcing algorithm for BLS is correct, secure, verifiable, and efficient.
\subsection{Correctness Analysis}
In this part, we prove the correctness of the outsourcing algorithm we proposed. As long as the cloud server executes in accordance with the algorithm, the correct result will be obtained by the client.

First, we prove that the result of the client's first recovery from the cloud server is correct. The proof is as follows:
\begin{equation}
\begin{aligned}
  {A}^{T}A&=({Q^{T}})^{-1}{A'}^{T}A'Q^{-1} \\
          &=({Q^{T}})^{-1}Q^{T}A^{T}AQQ^{-1}\\
          &=(({Q^{T}})^{-1}Q^{T})A^{T}A(QQ^{-1})\\
          &={A}^{T}A.
\end{aligned}
\end{equation}\
According to the above derivation process, we can prove that the client can recovery the result by Eq.(\ref{eq14}). Next we prove the correctness of the final result $R_{4}$ obtained from the cloud. The derivation process of $R_{4}$ is as follows:
\begin{equation}
\begin{aligned}
    R_{4}&=QR_{3}P\\
         &=QR_{2}^{-1}{A'}^{T}P\\
         &=QQ^{-1}R_{1}^{-1}(Q^{T})^{-1}Q^{T}A^{T}P^{T}P\\
         &=(QQ^{-1})R_{1}^{-1}((Q^{T})^{-1}Q^{T})A^{T}(P^{T}P)\\
         &=R_{1}^{-1}A^{T}\\
         &=\lim_{\lambda\rightarrow0}(\lambda I + A^{T}A)^{-1}A^{T}.
\end{aligned}
\end{equation}
Consequently, $R_{4}$ is the final calculation result. Finally we derived the verification process and proved that the verification result must be correct. The client verifies whether the calculation result is correct by checking whether the equation $R_{4}A \gamma= \gamma$ holds. If $R_{4}$ is the correct result, the derivation process is as follows:
\begin{equation}
\begin{aligned}
    R_{4}A \gamma&=\lim_{\lambda\rightarrow0}(\lambda I + A^{T}A)^{-1}A^{T}A\gamma&\\
                 &=A^{+}A\gamma\\
                 &=(A^{+}A)\gamma\\
                 &=\gamma.
\end{aligned}
\end{equation}

Based on the above derivation, we can conclude that as long as the cloud server honestly executes the algorithm, the correct calculation result can be obtained and the client can effectively verify the correctness of the calculation result.

\subsection{Security Analysis}
\textbf{Input Privacy}:
In our outsourcing algorithm, as shown in Eq.(\ref{eq11}), the original matrix $A$ is obscured into matrix $A' = PAQ$, and the second input data $R_{1}$ sent to cloud is obscured as Eq.(\ref{eq15}). All sensitive information of clients is hidden by $P$ and $Q$. As shown in Eq.(\ref{eq8}) and Eq.(\ref{eq10}), $P$ and $Q$ are generated by random permutation. The probability of cloud server getting $P$ is $\frac{1}{2^{n}n!}$, and the probability of getting $Q$ is $\frac{1}{n^{n}n!} $. Thus, even if the cloud has input matrices $A'$ and $R_{2}$, the probability of recovering the client's original data can be ignored.

\textbf{Output Privacy}:
In our outsourcing algorithm, the computation on the cloud server mainly involves matrix multiplication and matrix inversion. The cloud will return the results to the client twice. The first output of the cloud is shown in Eq.(\ref{eq13}). The key matrix $Q$ obscures the result ${A}^{T}A$ that client really wants to obtain. Without knowing the key matrix $Q$, the probability of the cloud server obtaining the original input matrix $A$ is $\frac{1}{n^{n}n!}$, which can be ignored.

The second output of the cloud is shown in Eq.(\ref{eq17}). The key matrices $P$ and $Q$ hide the pseudoinverse. According to Eq.(\ref{eq8}) and Eq. (\ref{eq10}), the probability of the cloud to infer the correct pseudoinverse result is $\frac{1}{2^{n}n!}{n^{n}n!}$. Thus, without knowing $P$ and $Q$, the cloud cannot recover the pseudoinverse from the output.

Based on the above analysis, we can conclude that our algorithm can ensure the input privacy and the output privacy.

\subsection{Verifiability Analysis}

The result verification methods are presented in Section IV-D. In this section, we prove that our proposed algorithm has robust cheating resistance.

We need to prove that client can detect cloud fraud with a probability of almost 1. The client sends the obscured input and computation tasks to the cloud server, and if the cloud server has fraudulent behavior, the client will receive an incorrect result. On the client, in order to verify the correctness of the result returned by the cloud server, client need to recover real result $R_{4}$ as shown in Eq.(\ref{eq18}). Then client check whether the result $R_{4}$ satisfies $R_{4}Ar=r$. According to the characteristics of the pseudoinverse matrix, we know that if the $R_{4}$ is wrong, then $R_{4}Ar=r$ must not hold. If $R_{4}$ satisfies $R_{4}Ar=r$, then $R_{4}$ must be the correct result that the client wants.

According to the above analysis, we can draw a conclusion that our outsourcing algorithm can detect the cheating behavior of cloud with a probability of almost 1, and at the same time, it can be ensured that all the correct results will pass the verification successfully.

\subsection{Efficiency Analysis}
In this section, we analyze the efficiency of our proposed algorithm in detail, including the computation cost of the client and the cloud. In addition, We also analyze the amount of calculation that our proposed algorithm saves for the client.

\begin{itemize}
\item \textbf{Client-Side Overhead:} To ensure the security of the algorithm, in our outsourcing algorithm, transformation, recovery and verification need to be performed on the client. We first analyze the key generation part of the transformation phase. Algorithm \ref{Algorithm1} shows the complete process of key generation, from which we can find that the time complexity of key generation is $O(n)$. Next, We analyzed the rest of the transformation phase. In the algorithm, we transform the original input twice. The first transformation is shown in Eq.(\ref{eq11}), and the second transformation is shown in Eq.(\ref{eq15}). Although the two transformation processes involve matrix multiplication, the time complexity only is $O( n^{2})$ because the key matrix is sparse. Thirdly, we perform the analysis on recovery operation. After the cloud returns the calculation result to the client, the client recovers it to get the pseudoinverse matrix $R_{4}$ as Eq.(\ref{eq18}).
According to the above, the time complexity of the decryption process is also $O(n^{2})$. Finally, we analyze the client's operation in the verification process. In the verification process, we make full use of the characteristics of the pseudoinverse matrix. As shown in Eq.(\ref{eq20}), the time complexity of verification also reduces to $O(n^{2})$ after introducing a random vector.

\item \textbf{Cloud-Side Overhead:} Overhead of cloud server is generated by computing the data sent by client according to the outsourcing algorithm. In our proposed algorithm, cloud operations include computing the product of two matrices and finding the inverse matrix. Because the obscured matrix sent to the cloud is the same size as the original matrix, there is no need for additional operations in the cloud. It implies that there is no additional computation introduced by our proposed algorithm. Therefore, in the outsourcing algorithm, the cost of calculating the pseudoinverse matrix by the cloud server is the same as the cost of calculating the pseudoinverse matrix by the client, and both are $O(n^{3})$.
\end{itemize}

\section{Experimental Analysis}
\begin{figure}[htb]
\begin{center}
\includegraphics[height=150pt,width=150pt]{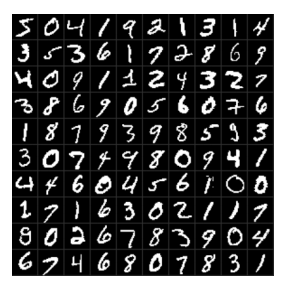}\\
\vspace{-3mm}
\begin{center}
\caption{Examples in MNIST data set\cite{2011Quickly}}
\label{MNIST} \vspace{-3mm}
\end{center}
\end{center}
\end{figure}

\begin{figure}[htb]
\begin{center}
\includegraphics[height=150pt,width=150pt]{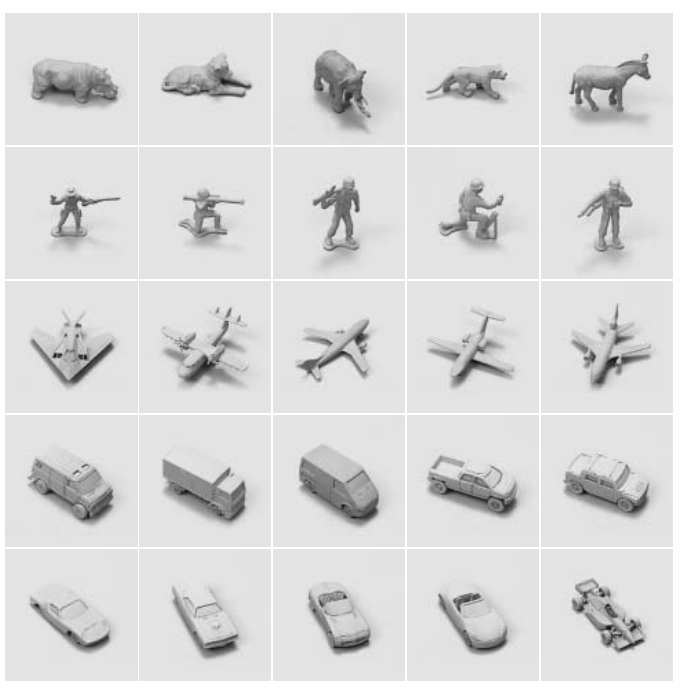}\\
\vspace{-3mm}
\begin{center}
\caption{Examples in NORB data set\cite{1315150}}
\label{NORB} \vspace{-3mm}
\end{center}
\end{center}
\end{figure}

In this section, in order to verify the effectiveness of our outsourcing algorithm, we conduct experiments and analyze the experimental results. In our experiment, we use a PC with an Intel-i5 2.4 GHz CPU and 8 GB of RAM to simulate a client and use a PC with an Xeon W-2123 3.6 GHz CPU and 32 GB of RAM to simulate a cloud server. All our experiments are implemented with the Pycharm platform using Python. In the experiment, we use the MNIST data set and the NORB data set as experimental samples. The MNIST data set contains 70000 handwritten digits, 60000 of which are used as the training set and 10000 as the testing set. These samples are from 250 different people, and each sample is represented by a 28 * 28 gray-scaled pixels, as shown in Fig.\ref{MNIST}. NORB is a 3D object image recognition data set. It contains 50 toy images in 5 general categories of quadrupeds, people, airplanes, trucks and cars. The shooting uses 2 cameras, 6 different lighting conditions, and 9 specific Shooting angles, 18 elevation angles. The training set includes 5 instances of each category, and the remaining 5 instances are the test set. The sample in NORB is shown in Fig.\ref{NORB}.

\subsection{Experimental Process}

\begin{figure*}[htbp]
\centering

\centering
\includegraphics[width=5  in]{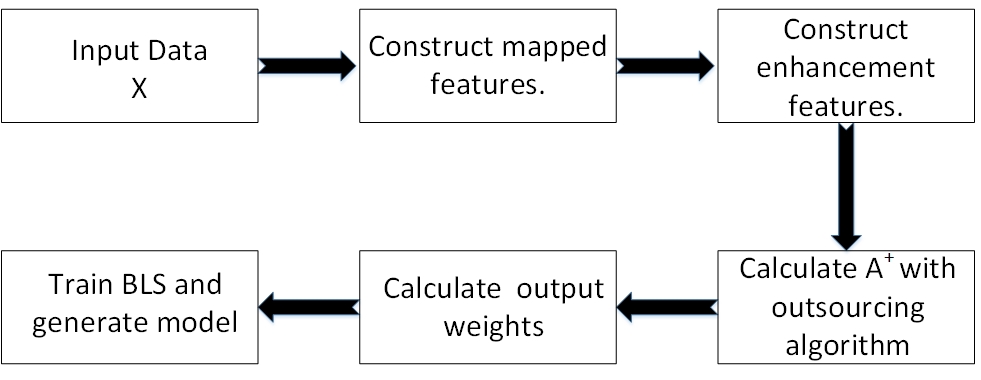}

\centering
\caption{Process of outsourcing BLS algorithm }\label{F5}

\end{figure*}

Fig.\ref{F5} shows the workflow of the outsourcing algorithm for BLS. After selecting data set $X$, the client constructs mapped feature nodes and enhancement nodes successively. The mapped feature nodes and enhancement nodes are horizontally spliced into matrix $A$, and $A$ is sent to the cloud server to outsource computing the pseudoinverse of $A$. After obtaining the pseudoinverse of $A$ by outsourcing algorithm, we calculate the global weight $W$ according to Eq.(\ref{eq3}). Next, we continue to train the BLS model. After training, we select some samples as the testing set, and generate the mapping feature nodes and enhancement nodes in the same way. Then we identify each test sample and calculate the accuracy of model identification.

\subsection{Evaluation Results}

\begin{figure*}[h!]
\centering
\scalebox{1}{
\subfigure[The time comparison on client]{
\begin{minipage}[t]{0.3\linewidth}
\centering
\includegraphics[width=1.5  in]{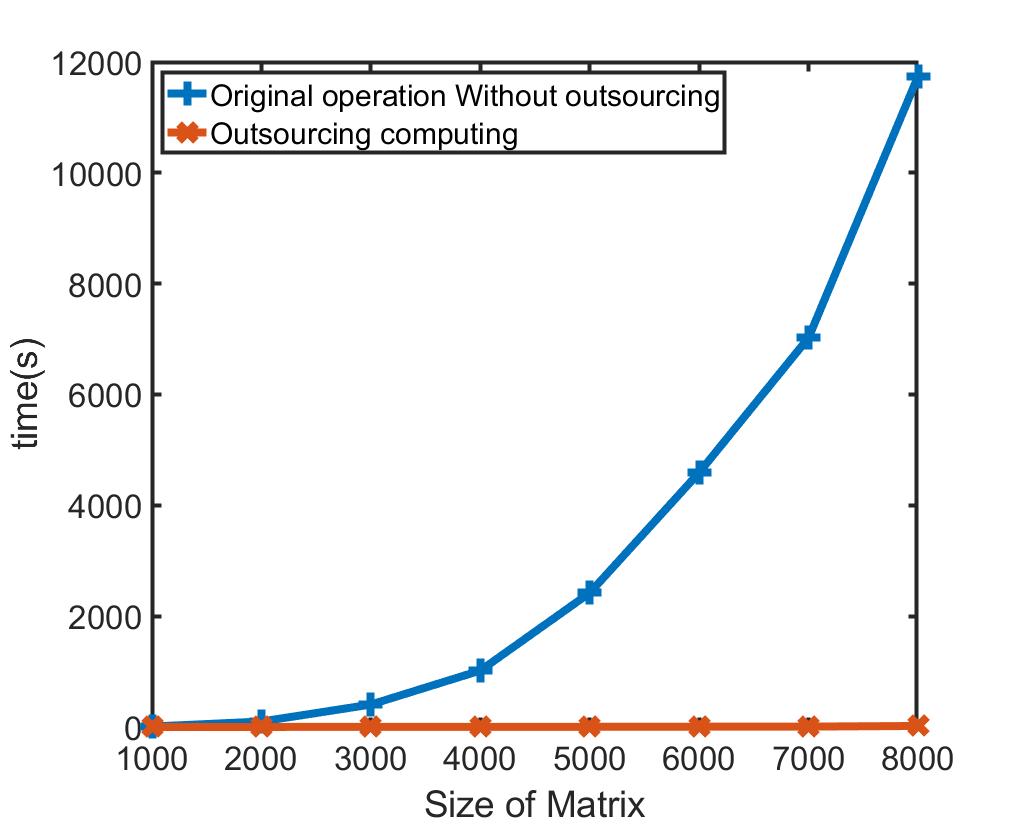}
\label{6.1}
\end{minipage}
}%
\subfigure[The time comparison in different phases ]{
\begin{minipage}[t]{0.3\linewidth}
\centering
\includegraphics[width=1.5  in]{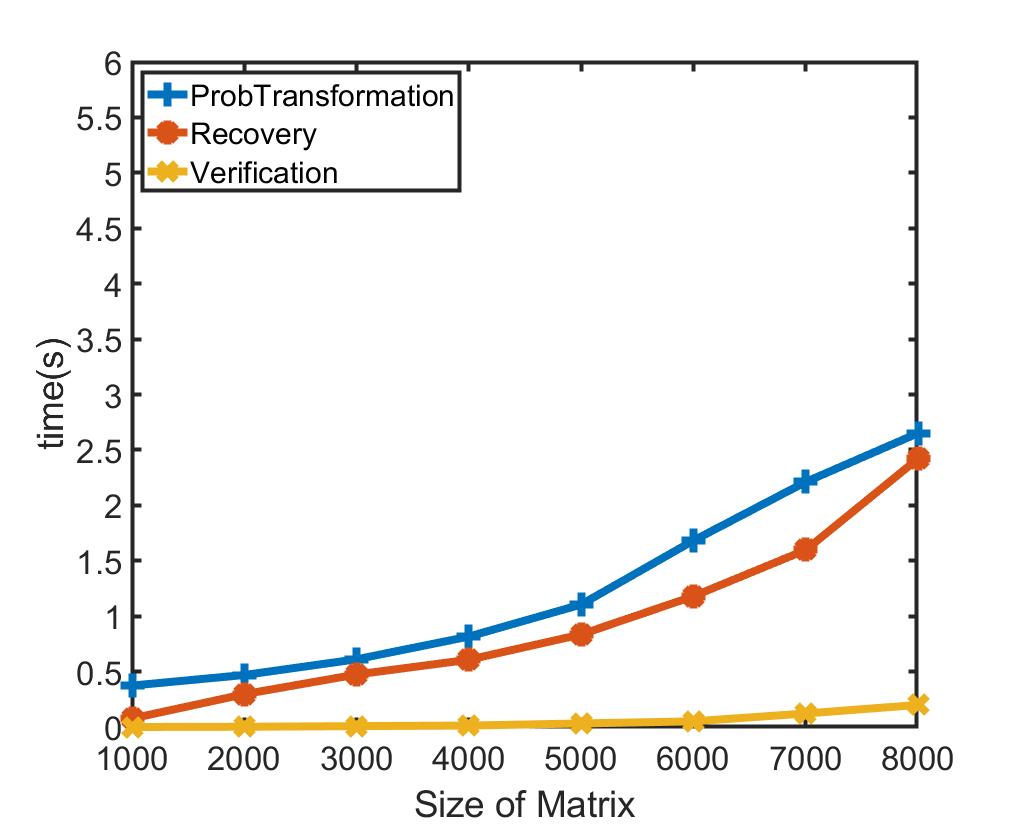}
\label{6.2}

\end{minipage}
}%
\subfigure[The time comparison on client and  cloud side ]{
\begin{minipage}[t]{0.3\linewidth}
\centering
\includegraphics[width=1.5  in]{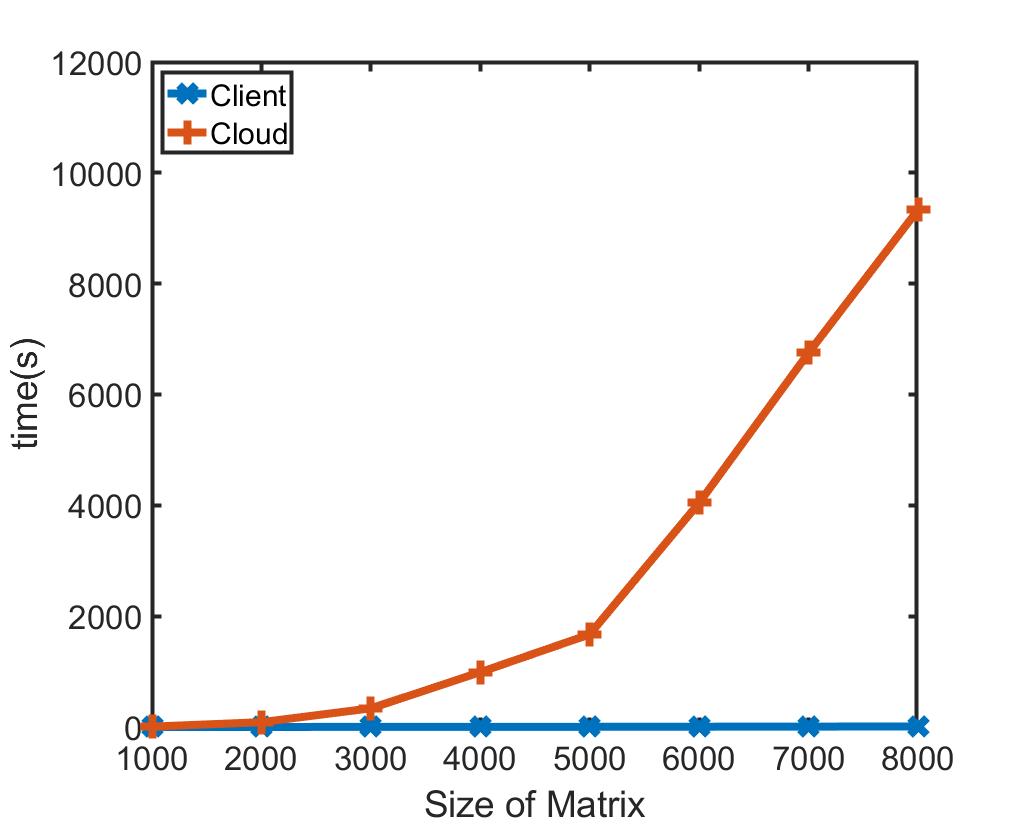}
\label{6.3}
\end{minipage}
}}%
\centering
\caption{Evaluation results for BLS computing outsourcing algorithm in MNIST }\label{6}

\end{figure*}

\begin{figure*}[h]
\centering
\scalebox{1}{
\subfigure[The time comparison on client]{
\begin{minipage}[t]{0.3\linewidth}
\centering
\includegraphics[width=1.5  in]{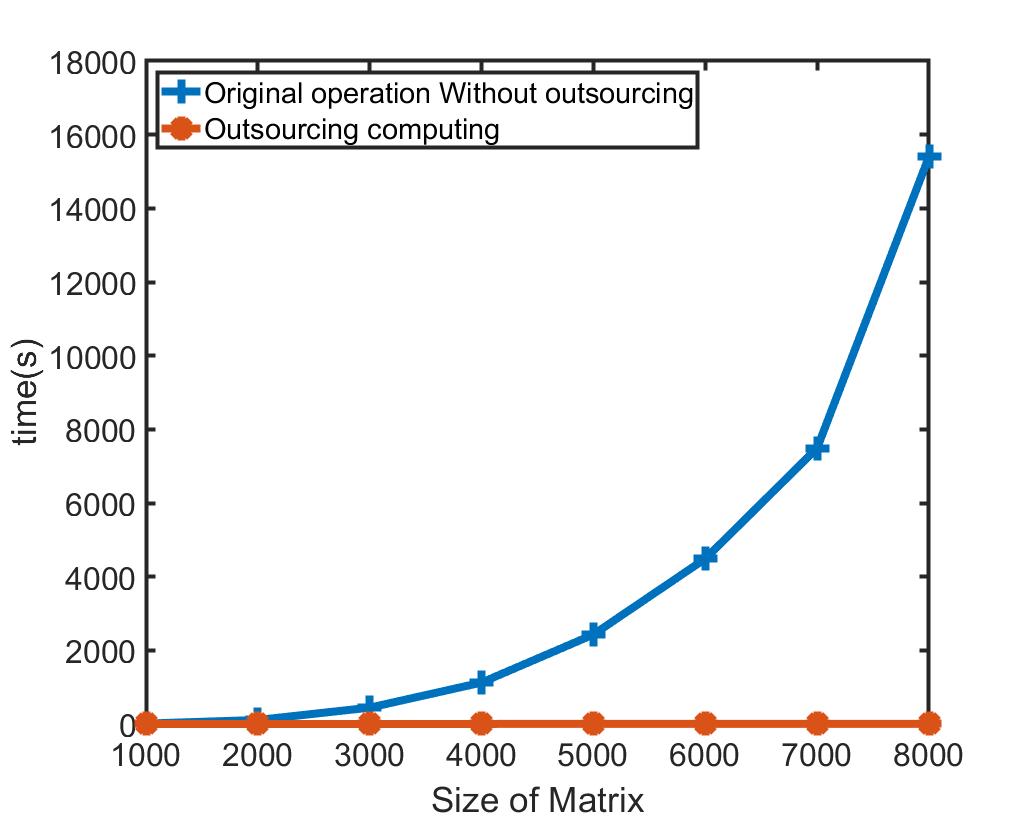}
\label{7.1}
\end{minipage}
}%
\subfigure[The time comparison in different phases ]{
\begin{minipage}[t]{0.3\linewidth}
\centering
\includegraphics[width=1.5  in]{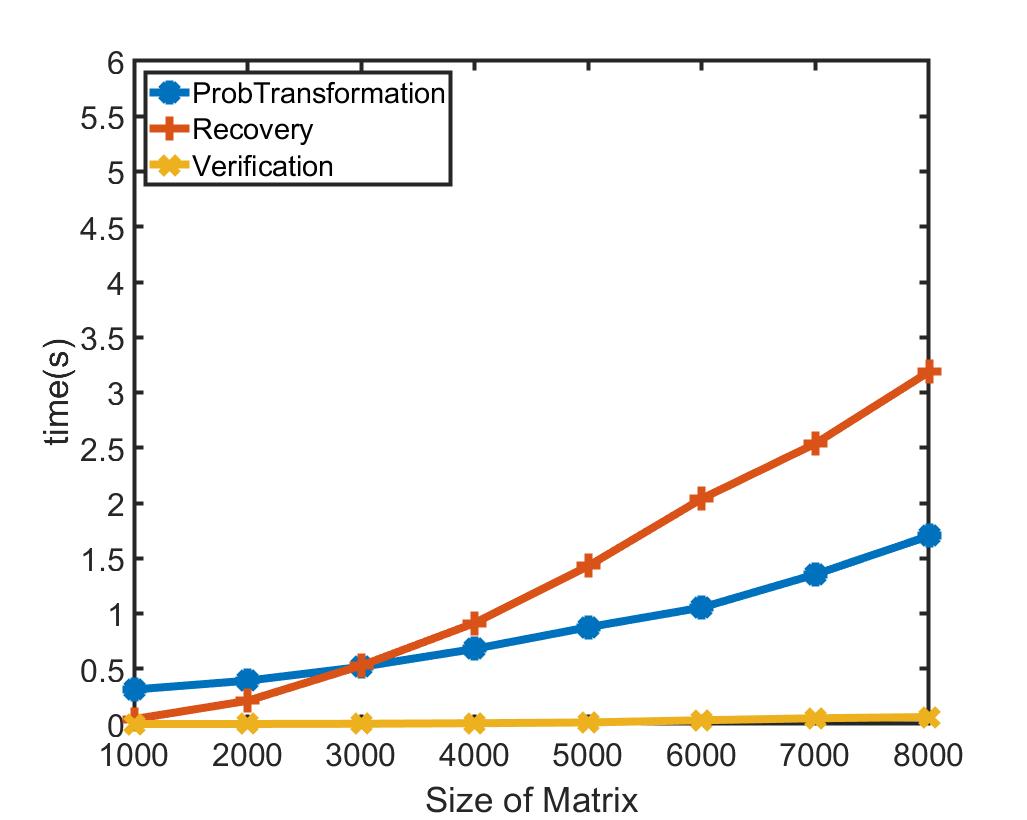}
\label{7.2}
\end{minipage}
}%
\subfigure[The time comparison on client and  cloud side ]{
\begin{minipage}[t]{0.3\linewidth}
\centering
\includegraphics[width=1.5  in]{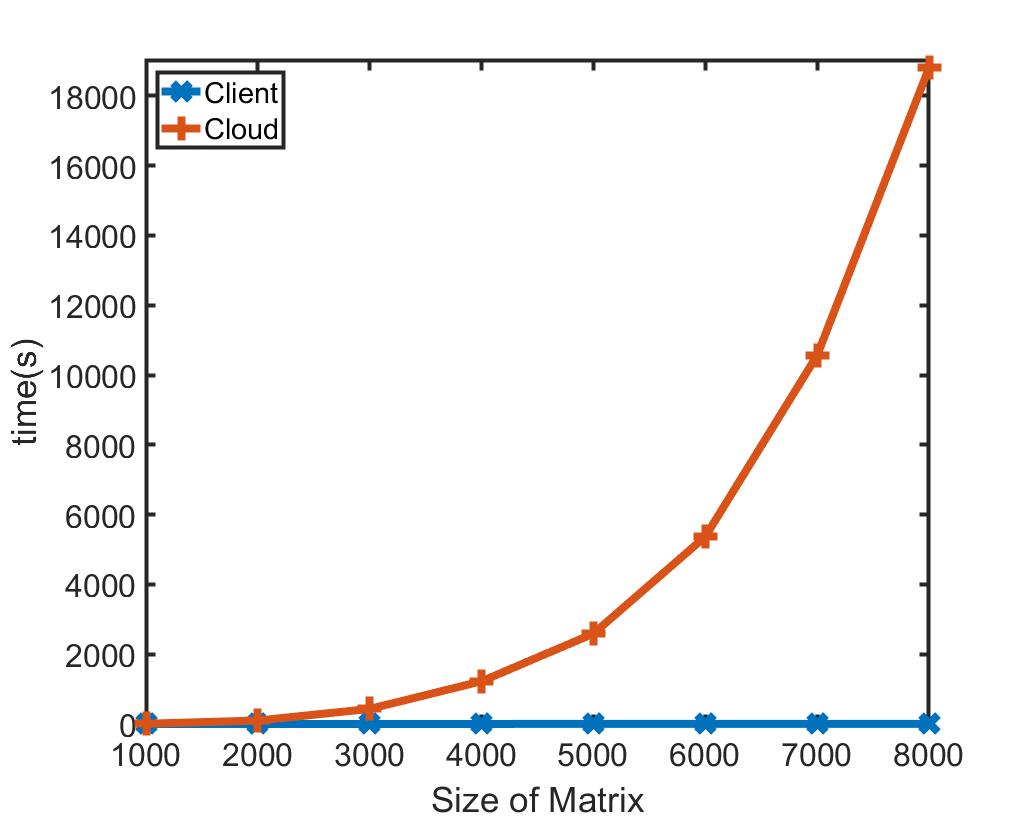}
\label{7.3}
\end{minipage}
}}%
\centering
\caption{Evaluation results for BLS computing outsourcing algorithm in NORB }\label{7}
\end{figure*}

We set the size range of input matrix $A$ of outsourcing computing part from $1000 \times 1000$ to $8000 \times 8000$. We evaluate the experimental results in terms of time efficiency and accuracy. Fig.\ref{6} and Fig.\ref{7} show the evaluation results of the experiment. Sub-figure \ref{6.1} and Sub-figure \ref{7.1} compare the time cost of the client who uses the outsourcing algorithm and the time cost of the client who does not use the outsourcing algorithm. We can find that when the size of the matrix is the same, there is a huge difference between the time cost to complete the training for the client who uses the outsourcing algorithm and who does not. Obviously, using our outsourcing algorithm can save the client a significant amount of time. The reason is that client who does not use outsourcing algorithm needs a lot of computing resources to complete the training process on the local side. After using the outsourcing algorithm, the complex calculation will be handed over to the cloud, and the client only needs to perform a small amount of simple calculation locally. Sub-figure \ref{6.2} and Sub-figure \ref{7.2} compare the time cost of the client among different phases of the outsourcing algorithm. From the figures, we can observe, the transformation and recovery phase is more time-consuming than the verification phase. The reason is that the client needs to conduct twice the transformation and recovery phase, and each time, the client needs to perform the sparse matrices multiplication twice. Meanwhile, the verification phase only needs one simple calculation as Eq.(\ref{eq20}). Sub-figure \ref{6.3} and Sub-figure \ref{7.3} compare the time cost of client and cloud server in the proposed outsourcing algorithm. The time cost of the cloud is much higher than that of the client, because the cloud needs to calculate the matrix multiplication, in which the time complexity is $O(n^{3})$, while the client only needs to perform transformation, recovery, and verification, in which the time complexity is only $O(n^{2})$.

\begin{table*}[h]

\centering
\scalebox{0.5}{
\centering
\renewcommand\arraystretch{3}{
\setlength{\tabcolsep}{5mm}{
\begin{tabular}{cccc}
\hline
Number of Feature Nodes & Number of Enhancement Nodes & Without Outsourcing Test Accuracy$(\%)$ & Outsourcing Test Accuracy$(\%)$ \\
\hline
200 & 800 & 55.76 & 55.76 \\
\hline
1000 & 1000 & 82.62 & 82.62 \\
\hline
1000 & 2000 & 88.19 & 88.19 \\
\hline
1000 & 3000 & 88.15 & 88.14 \\
\hline
1000 & 4000 & 88.75 & 88.41 \\
\hline
1000 & 5000 & 91.44 & 91.31 \\
\hline
1000 & 6000 & 91.39 & 91.39 \\
\hline
1000 & 7000 & 92.89 & 92.73 \\
\hline
\end{tabular}}}}\caption{Recognition accuracy of two algorithms in MNIST}\label{table1}
\end{table*}

\begin{table*}[h]

\centering
\scalebox{0.5}{
\renewcommand\arraystretch{3}{
\setlength{\tabcolsep}{5mm}{
\begin{tabular}{cccc}
\hline
Number of Feature Nodes & Number of Enhancement Nodes & Without Outsourcing Test Accuracy$(\%)$ & Outsourcing Test Accuracy$(\%)$ \\
\hline
200 & 800 & 60.77 & 60.77 \\
\hline
1000 & 1000 & 71.72 & 71.72 \\
\hline
1000 & 2000 & 81.77 & 77.31 \\
\hline
1000 & 3000 & 84.21 & 81.74 \\
\hline
1000 & 4000 & 88.75 & 84.16 \\
\hline
1000 & 5000 & 85.69 & 85.69 \\
\hline
1000 & 6000 & 84.93 & 84.94 \\
\hline
1000 & 7000 & 85.88 & 85.94 \\
\hline
\end{tabular}}}}\caption{Recognition accuracy of two algorithms in NORB}\label{table2}
\end{table*}
To evaluate the accuracy of our proposed outsourcing algorithm, we compare the recognition accuracy of the outsourcing algorithm with the original BLS algorithm. The comparison results are shown in Table \ref{table1} and Table \ref{table2}. From these two tables, we can find that the recognition accuracy of the two algorithms is slightly different. This is because in our outsourcing algorithm, in order to ensure that the client's data is not leaked by the cloud, we need to transformation or recovery the data every time the client communicates with the cloud server, the accuracy of the data after transform or recover may be slightly different, thus the accuracy of recognition is changed. The above experiments show that our algorithm greatly reduces the time cost of the client with only a small change in the accuracy rate, which is of great significance for the promotion of BLS.

\section{Related Work} \label{sec:rel}
In this section, we review the development of cloud computing and secure outsourcing computations. We also introduce the related researches of the  BLS.

\subsection{Cloud Computing}
Cloud computing is an Internet-based computing resource interaction mode, which integrates various services, applications and other resources\cite{Buyya2008Market,article}. Cloud computing is a service that can provide resources on-demand in a short period of time. The development of cloud computing greatly promotes the new application of traditional technology in the cloud computing platform \cite{article2,Vaquero2009A}. Barroso \emph{et al}. in \cite{BARROSO2003Web} proposed an architecture for search applications based on the distributed characteristics of cloud servers, which allows a single query to use multiple servers. Kienzler \emph{et al}. in \cite{Kienzler2012Large} proposed a large-scale data processing method leveraging the cloud server. This method provides an incremental data processing model, which makes full use of the powerful computing power of the cloud server, meanwhile hides the data transmission delay. Shen \emph{et al}. in \cite{7805146} proposed a distributed computing scheme using the cloud server. The scheme uses distributed collaboration to obtain the computing results, which makes full use of the computing resources, meanwhile ensures the privacy and correctness of the final solution. Robert L. Grossman \emph{et al}. in \cite{article3} proposed a mining method using high-performance cloud servers. This scheme uses the storage and computing services provided by the cloud server to archive, analyze, and mine large distributed data sets.

\subsection{Secure Outsourcing Computations}

Although cloud computing has brought many conveniences to people, it also has brought some new challenges \cite{Atallah2002Secure13,Chen2014Highly22}. Secure outsourcing computations have attracted attention from researches. In \cite{David2008Private33,Atallah2002Secure34}, some secure outsourcing solutions for specific computing were proposed. However, these solutions are based on cloud fully trusted and cannot verify the results. In the current cloud computing environment, cloud servers are monopolized in private enterprises, and therefore can not be fully trusted.

To solve the above problems, the research direction of outsourcing computing has turned to the design of a secure outsourcing algorithm under a malicious cloud model. In recent years, the secure outsourcing computations of matrix operations under the malicious cloud model has attracted more and more attention from researchers. Chen \emph{et al}. in \cite{Chen2015New37} proposed an algorithm for solving large-scale linear equations. The algorithm makes full use of the characteristics of the sparse matrix and realizes the outsourcing calculation of large-scale linear equations under the complete malicious cloud model.
Luo \emph{et al}. in \cite{Luo2017SecFact40} proposed a secure outsourcing algorithm for large-scale QR decomposition and LU decomposition. The algorithm uses the rich computing resources of the cloud, so that client can efficiently perform large-scale QR and LU factorization. Zhang \emph{et al}. in \cite{2020Practical} proposed an outsourcing scheme that can securely use cloud servers to solve the quadratic congruences problem in the Internet of Things. This algorithm solves the problem that resource-constrained IoT devices cannot calculate the quadratic congruences. While improving the computing efficiency of IoT devices, this algorithm also ensures data privacy. Wang \emph{et al}. in \cite{Wang2011Secure38}  proposed a secure outsourcing solution for linear programming problems. In the algorithm, the original linear programming problem is transformed into a random problem. Salinas \emph{et al}. in \cite{Salinas2015Efficient22} proposed a secure outsourcing algorithm to efficiently solving large-scale sparse linear systems of equations. The algorithm not only reduces the time required for client calculations, but also reduces the memory I/O operations by the client. Zhang \emph{et al}. in \cite{2020Zhang} proposed a secure edge computing framework for the matrix multiplication, which allows resource-constrained IoT devices to efficiently and safely complete complex matrix calculations by using edge servers.

\subsection{BLS Development History}
Broad Learning System (BLS) is an efficient incremental learning algorithm proposed by Chen \emph{et al} \cite{C6}. It solves the time-consuming problem of repeated training in some cases by incremental learning. However, the accuracy of the original BLS model is not so satisfactory in some large-scale data sets. Hence, on the basis of the original algorithm of BLS, many researchers have improved the structure of BLS. Liu \emph{et al} in \cite{Liu2017Broad} proposed an improved BLS  based on K-means clustering algorithm. Compared with the original system, it achieves satisfactory performance on more complex data sets. Zheng \emph{et al}. in \cite{9147058} proposed a system based on the maximum correntropy criterion and BLS. To obtain a more robust BLS, they adopt the maximum entropy criterion to train the output weight of the BLS. With the continuous improvement of BLS, some researchers have used BLS to achieve impressive performance in image processing and other fields. Lin \emph{et al}. in \cite{af10} proposed a flexible approach for human activity recognition based on BLS, which improves the model training speed and prediction accuracy by introducing BLS. Kong \emph{et al}. in \cite{article1} proposed an image classification method based on semi-supervised BLS. This method merges the category probability structure into the generalized learning model to obtain a semi-supervised generalized learning version and introduces BLS into the hyperspectral imagery (HSI) classification, thereby improving efficiency. After the above improvements, BLS becomes more and more powerful. However, no one has proposed an outsourcing algorithm for BLS.

\section{Conclusion}

In this paper, we design a secure and efficient outsourcing algorithm for the BLS, in which the time complexity on the local client is reduced from $O(n^{3})$ to $O(n^{2})$. As a consequence, resource-constrained devices can efficiently accomplish the complex training process of the BLS. In the outsourcing process, our algorithm ensures that the original inputs and output are kept secret to the cloud server. And the client can verify the results returned by cloud with a probability of almost 1. Also, we verify the efficiency of the algorithm through experiments. With our proposed algorithm, the BLS can be applied in a broader range of applications regardless of the computation power of end devices, which promotes the popularization of the BLS.

\section{Acknowledgement}

This research is supported by National Natural Science Foundation of China (61572267), National Development Foundation of Cryptography (MMJJ20170118), the Joint Found of the National Natural Science Foundation of China (U1905211), Key Research and Development Project of Shandong Province (2019GGX101051), K. C. Wong Education Foundation, Natural Science Basic Research Plan in Shaanxi Province of China (2019JQ-124).

\section*{References}

\bibliography{mybibfile}
\end{spacing}
\end{document}